Original Paper

# Social Media Polarization and Echo Chambers in the Context of COVID-19: Case Study

Julie Jiang[1,2], BSc; Xiang Ren[2], PhD; Emilio Ferrara[1,2,3], PhD

[1]Information Sciences Institute, University of Southern California, Marina Del Rey, CA, United States
[2]Department of Computer Science, Viterbi School of Engineering, University of Southern California, Los Angeles, CA, United States
[3]Annenberg School of Communication, University of Southern California, Los Angeles, CA, United States

**Corresponding Author:**
Emilio Ferrara, PhD
Information Sciences Institute
University of Southern California
4676 Admiralty Way #1001
Marina Del Rey, CA, 90292
United States
Phone: 1 310 448 8661
Email: ferrarae@isi.edu

## Abstract

**Background:** Social media chatter in 2020 has been largely dominated by the COVID-19 pandemic. Existing research shows that COVID-19 discourse is highly politicized, with political preferences linked to beliefs and disbeliefs about the virus. As it happens with topics that become politicized, people may fall into echo chambers, which is the idea that one is only presented with information they already agree with, thereby reinforcing one's confirmation bias. Understanding the relationship between information dissemination and political preference is crucial for effective public health communication.

**Objective:** We aimed to study the extent of polarization and examine the structure of echo chambers related to COVID-19 discourse on Twitter in the United States.

**Methods:** First, we presented Retweet-BERT, a scalable and highly accurate model for estimating user polarity by leveraging language features and network structures. Then, by analyzing the user polarity predicted by Retweet-BERT, we provided new insights into the characterization of partisan users.

**Results:** We observed that right-leaning users were noticeably more vocal and active in the production and consumption of COVID-19 information. We also found that most of the highly influential users were partisan, which may contribute to further polarization. Importantly, while echo chambers exist in both the right- and left-leaning communities, the right-leaning community was by far more densely connected within their echo chamber and isolated from the rest.

**Conclusions:** We provided empirical evidence that political echo chambers are prevalent, especially in the right-leaning community, which can exacerbate the exposure to information in line with pre-existing users' views. Our findings have broader implications in developing effective public health campaigns and promoting the circulation of factual information online.

*(JMIRx Med 2021;2(3):e29570)*   doi: 10.2196/29570

XSL•FO
RenderX

## Introduction

### Background

As the unprecedented COVID-19 pandemic continues to put millions of people at home in isolation, online communication, especially on social media, is seeing a staggering uptick in engagement [1]. Prior research has shown that COVID-19 has become a highly politicized subject matter, with political preferences linked to beliefs (or disbeliefs) about the virus [2,3], support for safe practices [4], and willingness to return to activities [5]. As the United States was simultaneously undergoing one of the largest political events—the 2020 presidential election— public health policies may have been undermined by those who disagree politically with health officials and prominent government leaders. As it happens with topics that become politicized, people may fall into *echo chambers*—the idea that one is only presented with information they already agree with, thereby reinforcing one's confirmation bias [6,7].

Social media platforms have been criticized for enhancing political echo chambers and driving political polarization [8-10]. In part, this is due to a conscious decision made by users when choosing who or what to follow, selectively exposing themselves to content they already agree with [6]. This may also be a consequence of the algorithms social media platforms use to attract users [9]. Numerous studies have shown that echo chambers are prevalent on Twitter [7,8,11-13]; however, most past works are done on topics that are political in nature. In the case of COVID-19, the risks of political polarization and echo chambers can have dire consequences in politicizing a topic that is originally of public health. The lack of diversity in multiperspective and evidence-based information can present serious consequences for society by fueling the spread of misinformation [14-16]. For instance, prior research revealed that conservative users push narratives contradicting public health experts (eg, antimask) and misinformation (eg, voter fraud) [17]. Another study showed that the consumption of conservative media is linked to an increase in conspiracy beliefs [18]. Understanding the degree of polarization and the extent of echo chambers can help policymakers and public health officials effectively relay accurate information and debunk misinformation to the public.

### Research Questions

In this paper, we focused on the issue of COVID-19 and presented a large-scale empirical analysis on the prevalence of echo chambers and the effect of polarization on social media. Our research was guided by the following research questions (RQs) surrounding COVID-19 discussions on Twitter:

- RQ1: What are the roles of partisan users on social media in spreading COVID-19 information? How polarized are the most influential users?
- RQ2: Do echo chambers exist? And yes, what are the echo chambers and how do they compare?

The technical challenge for addressing these questions is posed by the need to build a scalable and reliable method to estimate user political leanings. To this end, we proposed Retweet-BERT, an end-to-end model that estimates user polarity from their profiles and retweets on a spectrum from left to right leaning.

## Methods

### Data

We used a large COVID-19 Twitter data set collected by Chen et al [19], containing data from January 21 to July 31, 2020 (v2.7). All tweets collected contain keywords relevant to COVID-19. The tweets can be an original tweet, retweets, quoted tweets (retweets with comments), or replies. Each tweet also contains the user's profile description, the number of followers they have, and the user-provided location. Some users are verified, meaning they are authenticated by Twitter in the interest of the public, reducing the chance that they are fake or bot accounts [20]. All users can optionally fill in their profile descriptions, which can include personal descriptors (eg, "Dog-lover," "Senator," "Best-selling author") and the political party or activism they support (eg, "Republican," "#BLM").

### Interaction Networks

The retweet network $G_R=(V,E)$ was modeled as a weighted, directed graph. Each user $u \in V$ is a node in the graph, each edge $(u,v) \in E$ indicates that user $u$ has retweeted from user $v$, and the weight of an edge $w(u,v)$ represents the number of retweets. We used the terms retweet interaction and edges of the retweet network interchangeably. Similarly, we constructed the mention network $G_M$, where the edges are mentions instead of retweets. A user can be mentioned through retweets, quoted tweets, replies, or otherwise directly mentioned in any tweet.

### Data Preprocessing

We restricted our attention to users who are likely located in the United States, as determined by their self-provided location [4]. Following Garimella et al [21], we only retained edges in the retweet network with weights of at least 2. Since retweets often imply endorsement [22], a user retweeting another user more than once would imply stronger endorsement and produce more reliable results. As our analyses depend on user profiles, we removed users with no profile data. We also removed users with degrees less than 10 (in- or out-degrees) in the retweet network, as these are mostly inactive Twitter users. To remove biases from potential bots infiltrating the data set [23], we calculated bot scores using the methodology of Davis et al [24], which estimates a score from 0 (likely human) to 1 (likely bots), and removed the top 10% of users by bot scores as suggested by Ferrara [23].

Our final data set contained 232,000 users with 1.4 million retweet interactions among them. The average degree of the retweet network was 6.15. For the same set of users in the mention network, there were 10 million mention interactions, with an average degree of 46.19. Around 18,000, or approximately 8% of all, users were verified.

### Estimating User Polarity

This section describes our proposed method to estimate the polarity of users in a spectrum from left to right. We first surveyed related work and used weak-supervision to detect two polarized groups of users, which we treated as seed users. Then, we explored various models to predict the political leaning of users. Finally, these models were evaluated on labeled data using 5-fold cross-validation and the best model was applied to the remaining users to obtain their polarity scores.

### Related Work

#### Representation Learning on Twitter

Analysis of Twitter data takes the form of two, often combined, approaches, namely content-based and network-based. In content-based approaches, users are characterized by the account metadata, hashtags, tweet content, and other language-related features extracted from their profiles [25-27]. In network-based approaches, users are represented in the retweet network or the mention network, both being directed networks where edges indicate the flow of communication [8,28]. The use of user-follower networks is rare due to the time-consuming nature of its data collection [29].

Both approaches can benefit from recent advances in representation learning, and specifically embedding methods. Techniques like word embedding [30], or more recently transformers [31], have been shown to improve sentiment analysis on tweets [32] and tweet topic classification [33]. These models generate a vector representation of text so that semantically similar words and texts share similar representations. The concept of word embeddings can also be applied to networks, where node presentations embody their homophily and structural similarity [34]. Network embedding can aid user-type detection. For instance, Ribeiro et al [35] used representation learning on both the retweet network structure and the tweet content to detect hateful users. Xiao et al [36] used network representations to classify users in a politically centered network. In this work, we proposed a new strategy based on combining content and network embedding for user polarity detection.

#### Ideology Detection

The ability to detect user ideology is of interest to many researchers, for example, to enable studies of political preference. Most methods are rooted in the observation that people sharing similar political beliefs are often situated in tightly knit communities [8]. Earlier methods (eg, Conover et al [8]) classified users' political leanings based on the hashtag they used. The same challenge has been tackled with label propagation, with users who have linked left-winged or right-winged media outlets in their tweets as seed users [26,27]. Barberá et al [7] proposed a latent space model to estimate the polarity of users, assuming that users tend to follow politicians who share similar ideological stances. Darwish et al [37] developed an unsupervised approach to cluster users who share similar political stances based on their hashtags, retweet texts, and retweet accounts. Word embeddings have also been applied to user tweets to generate clusters of topics, which helps inform the political leaning of users [38]. Recently, Xiao et al [36] formulated a multirelational network to detect binary ideological labels. Our proposed method stands out because it (1) combines both language and network features for a more comprehensive estimation of ideology, and (2) is scalable and can be trained within a limited time with limited labeled data.

#### Pseudo Label Generation

We used two weakly supervised strategies to find the pseudo labels of political leanings for a subset of users (ie, seed users). For the first method, we gathered the top 50 most-used hashtags in user profiles and annotated them as left- or right-leaning depending on what political party or candidate they support (or oppose). Of these hashtags (uncased), 17 were classified as left-leaning (eg, #TheResistance, #VoteBlue) and 12 as right-leaning (eg, #MAGA, #KAG). Users were labeled as left-leaning or right-leaning if their profile contains more left-leaning or right-leaning hashtags, respectively. We did not consider hashtags used in tweets, for the reason that hashtags in tweets can be used to inject opposing content into the feed of other users [8]. Instead, in line with Badawy et al [26] and Addawood et al [27], we assume that hashtags appearing in user profiles would more accurately capture true political affiliation.

An alternative method makes use of the media outlets mentioned in users' tweets through mentions or retweets [39-41]. Similar to Ferrara et al [41], we identified 29 prominent media outlets on Twitter. Each media outlet has its media bias scored by the nonpartisan media watchdog AllSides.com on a scale of 1 to 5 (left, center-left, neutral, center-right, right). An endorsement from a user was defined as either an explicit retweet from a media's official Twitter account or a mention of a link from the media's website. Given a user who has given at least two endorsements, we calculated their media bias score from the average of the scores of their media outlets. A user was considered left-leaning if their media bias score was equal to or below 2 or right-leaning if above 4.

Using a combination of the profile hashtag method and the media outlet method, we categorized 79,370 (34% of all) users as either left- or right-leaning. In case of any disagreements between the two detection methods, we deferred to the first one (the hashtag-based method). We referred to these users as seed users for political leaning estimation. A total of 59,832, or 75% of all, seed users were left-leaning, compared to 19,538 who were right-leaning, consistent with previous research which revealed that there are more liberal users on Twitter [42].

#### Polarity Estimation Models

To predict user political leanings, we explored several representation learning methods based on the users' profile description and/or their retweet interactions. We provided an overview of natural language processing techniques to extract information from profile descriptions, as well as network embedding techniques to extract information from retweet interactions. We then proposed a new model that includes both components. All models were evaluated on the binary classification task of predicting (pseudo label) political leanings for the subset of seed users.

In the following two subsections, we describe various ways to get word embeddings, sentence (ie, profile) embeddings, and node embeddings. An embedding is a low-dimensional, vectorized representation of the word, sentence, or node relative to other inputs of the same kind. Embeddings capture the semantic (for language) or structural (for network) similarity of the inputs. Embeddings can be pretrained and transferred across data sets or tasks. Once trained, every word, sentence, or node can be mapped to a continuous vector embedding, where semantically similar words or sentences or structurally similar nodes share similar embeddings with each other.

**Language-Based Methods**

*Word Embeddings*

Word2Vec [30] and GloVe [43] are word embedding methods that learn word associations from a large corpus of text without supervision. Word2Vec considers a word and its surrounding words as the context in a sentence, while GloVe considers the global word-word co-occurrence matrix. Once trained, both models produce embeddings that capture the semantic similarity between words.

As baselines, we used pretrained Word2Vec and GloVe word embeddings from Gensim [44]. We formed profile embeddings by averaging the word embeddings of each word in the profile description. We fit a logistic regression model on the profile embeddings for the classification task.

*Transformers*

Transformers such as BERT [31], RoBERTa [45], and DistilBERT [46] are pretrained language models that have led to significant performance gains across many natural language processing tasks. Unlike word embeddings, transformers can disambiguate words with different meanings under different contexts. Transformers are deep learning models that are trained to understand sequential texts by way of predicting missing tokens (words) in the text and/or predicting the next sentence. They are also designed to easily adapt to various downstream tasks by fine-tuning the output layers.

There are a few ways to adapt transformers for profile classification. Transformers, which are already pretrained, can be directly applied to each individual profile. The outputs of a transformer include an initial token embedding (eg, [CLS] for BERT, <s> for RoBERTa) of the profile description as well as contextualized word embeddings for each token of the profile. One way to use transformers for classification is to *average* the output embeddings of each word in the profile, followed by a logistic regression model. The other, more time-consuming method is to *fine-tune* the *head* of the transformer through the initial token embedding by adding a set of deep-learning layers designed for classification. We used the sequence classification head published with *HuggingFace*'s open-sourced transformers library [47], which adds a linear dense layer on top of the pooled output of the initial token embedding of the transformers. This classification head outputs a single value between 0 and 1 using a sigmoid activation function.

*S-BERT*

Transformers in and of themselves are not suitable for large-scale sentence-based tasks. To remedy this, Reimers and Gurevych [48] proposed Sentence Transformers (S-BERT), which consists of Siamese and triplet networks to produce semantically meaningful sentence embeddings. S-BERT outperforms naive transformer-based methods for semantic textual similarity tasks, while massively reducing the time complexity. During training, S-BERT takes two sentences in parallel through an identical transformer (Siamese), adds a pooling operation to their outputs, and learns to predict predefined sentence pair objectives, such as measuring the similarity between the two sentences.

Using S-BERT models pretrained for semantic textual similarity, we retrieved embeddings for every profile. The profile embeddings were fit with a logistic regression model for classification.

**Network-Based Methods**

Similar to how word or sentence embeddings can be generated text, we can generate node embeddings for nodes in a network. Such node embeddings can capture network structure similarities and homophily. One network embedding model is node2vec [49], which learns node embeddings from random walks over the network. An important drawback of node2vec is that it cannot be used on isolated nodes. GraphSAGE [50] is another network embedding method that also utilizes node attributes and is inductive, meaning it can be applied to isolated nodes. We can use any of the aforementioned profile embeddings retrieved from any language models as the node attributes.

Another popular network-based method for node classification is label propagation, which deterministically propagates labels from seed users in the network. Label propagation also cannot predict for isolated nodes.

*Proposed Method: Retweet-BERT*

Inspired by S-BERT [48], we propose Retweet-BERT (visualized in Figure 1), a sentence embedding model that incorporates the retweet network. We based our model on the assumption that users who retweet each other are more likely to share similar ideologies. As such, the intuition of our model is to make profile embeddings more similar for users who retweet each other. Specifically, using any of the aforementioned models that can produce sentence-level embeddings, let $s_i$ denote the profile embedding for user $i$. For every positive retweet interaction from user $i$ to $j$ (ie, $(i,j) \in E$), we optimized the objective:

$$\sum_{k \in V, (i,k) \notin E} \max(\|s_i - s_j\| - \|s_i - s_k\| + \epsilon, 0) \quad (1)$$

where $\|\cdot\|$ is a distance metric and $\epsilon$ is a margin hyperparameter. We followed the default configuration of S-BERT, which uses the Euclidean distance and $\epsilon = 1$.

To optimize the training procedure, we used two negative sampling strategies. The first was negative sampling (one-neg), in which we randomly sampled one other node $k$ for every anchor node in each iteration [30]. For simplicity, we assumed all nodes are uniformly distributed. The second was multiple negative sampling (mult-neg), in which the negative examples are all of the other examples in the same batch [51]. For instance, if the batch of positive examples are $[(s_{i1}, s_{j1}), (s_{i2}, s_{j2}), ..., (s_{in}, s_{jn})]$, then the negative examples for pair at index $k$ are $(s_{ik}, s_{jk})$ are all the $\{s_{jk'}\}$ for $k' \in [1, n]$ and $k' \neq k$.

It is worth noting that Retweet-BERT disregards the directionality of the network and only considers the immediate neighbors of all nodes. In practice, however, we find that this model balances the trade-off between training complexity and testing performance. Building on the convenience of S-BERT for sentence embeddings, we used the aforementioned S-BERT models pretrained for semantic textual similarity as the basis for fine-tuning.

**Figure 1.** Illustration of the proposed Retweet-BERT. We first fine-tuned it on the retweet network (left) using a Siamese network structure, where the two BERT networks share weights. We then trained a denser layer on top to predict polarity (right).

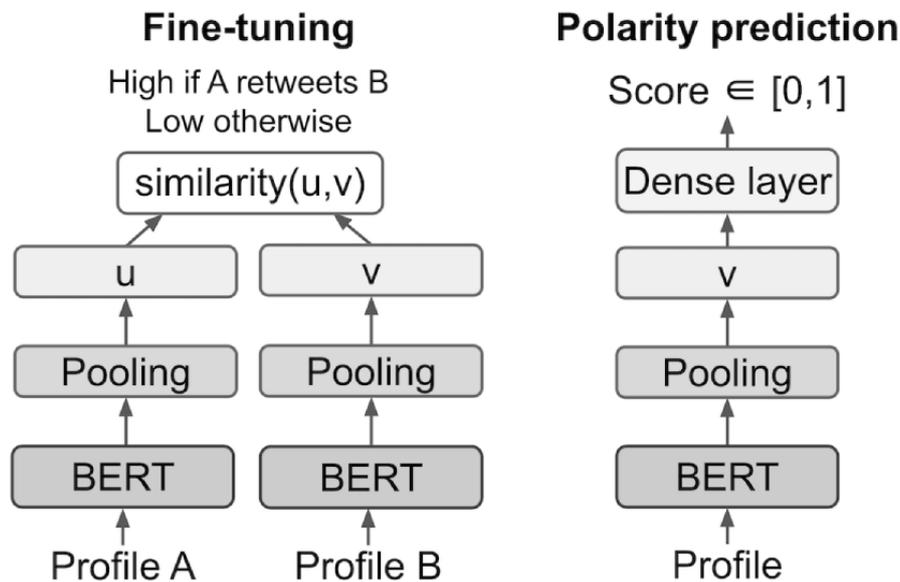

*Polarity Estimation Results*

We included an overview of the experiment results in Multimedia Appendix 1. Our proposed model, Retweet-BERT, achieves the best result with a BERT base model trained with the multiple negatives training strategy. It attains 96% cross-validated AUC (area under the receiver operating characteristic curve), which is a common metric for use in measuring binary classification in unbalanced classes. Previously, we also conducted an in-depth evaluation of our model (Jiang et al, unpublished work). We trained Retweet-BERT on all of the seed users with political leaning pseudo labels and inferred polarity scores for the rest of the users, ranging from 0 (far-left) to 1 (far-right). These scores will be referred to as *polarity scores*. Since there were more left-leaning seed users, the polarity scores were naturally skewed toward 0 (left). Therefore, similar to previous work [23,26,28], we binned users by evenly distributed deciles of the polarity scores, with each decile containing exactly 10% of all users.

## Results

### The Roles of Partisan Users

We first examined the characteristics of extremely polarized users, defined as the users in the bottom (left-leaning/far-left) or top (right-leaning/far-right) 20% of the polarity scores. As a point of comparison, we also included neutral users who were in the middle 20% of the polarity scores. Considering various aspects of user tweeting behaviors, we characterized the Twitter user roles as follows:

1. Information creators: those who create original content and are usually the source of new information.
2. Information broadcasters: those who foster the distribution of existing content, such as through retweeting other people and promoting the visibility of other's content.
3. Information distributors: those whose contents are likely to be seen by many people, either through passive consumption by their followers or through broadcasting (retweeting) by others.

According to these definitions, a user can be all of these or none of these at the same time. In Figure 2, we plot several Twitter statistics regarding the polarized and neutral users, disaggregated by their verification status.

Compared to unverified users, verified users were more likely to be information creators. This is unsurprising, given that verified users can only be verified if they demonstrate they are of public interest and noteworthy. Comparatively, left-leaning verified users had the smallest fraction of original posts. However, this was reversed for unverified users, with unverified left-leaning users having the highest fraction of original content and unverified right-leaning users having little to no original content. We noted that this may be related to the distribution of bot scores. If bots infiltrated users of different partisanship equally, we expect to find similar distributions of bot scores across all users. However, Figure 2B reveals that right-leaning users scored significantly higher on the bot scale. Since bots retweet significantly more than normal users [52], we cannot rule out the possibility that right-leaning bots were confounding the analysis, even though those scoring the highest on the bot scale have already been removed from the data set.

Unverified right-leaning users, in comparison with their left-leaning counterparts, were more likely to be information broadcasters as they had the highest out-degree distribution (Figure 2C). As out-degree measures the number of people a user retweets from, a user with a high out-degree plays a critical role in information broadcasting. The fact that they also had very little original content (Figure 2A) further suggests that unverified right-leaning users primarily retweeted from others.

Finally, all right-leaning users functioned as information distributors regardless of their verification status. Their tweets were much more likely to be shared and consumed by others. Their high in-degree distribution indicates they got retweeted more often (Figure 2D), and the higher number of followers they have indicates that their posts were likely seen by more people (Figure 2E).

As right-leaning users played larger roles in both the broadcasting and distributing of information, we questioned if these users formed a political echo chamber, wherein right-leaning users retweet frequently from, but only from, users who are also right-leaning. As shown later in the paper, we did indeed find evidence that right-leaning users form a strong echo chamber.

**Figure 2.** Data set statistics of left-leaning (bottom 20%), neutral (middle 20%), and right-leaning (top 20%) users, partitioned by their verification status. The degree distributions are taken from the retweet network. All triplets of distributions (left-leaning, neutral, and right-leaning) are significantly different using a one-way ANOVA (analysis of variance) test (*P*<.001).

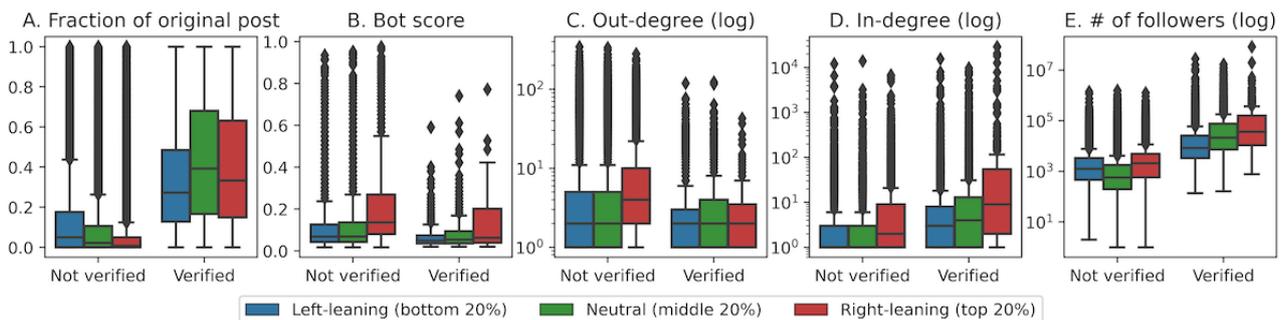

## The Polarity of Influencers

The above characterizes the Twitter activities of users who are extremely left- or right-biased. However, the majority of the social influence is controlled by a few key individuals [53-55]. In this section, we considered five measures of social influence: verification status, number of followers, number of retweets, number of mentions, and PageRank in the retweet network [56]. A user is considered influential if they are in the top 5% of all people according to the measure of influence. Figure 3 reveals the proportion of users in each decile of the polarity score that is influential. We showed that consistent with all of the influence measures above, partisan users are more likely to be influential.

The verification status is correlated with partisan bias, with the proportion of verified users decreasing linearly as we move from the most left- to the most right-leaning deciles of users (Figure 3A). Of the total, 15% of users in the first and second deciles, which are most liberal, were verified, compared to less than 1% of users in the extremely conservative 10th decile. As verified accounts generally mark the legitimacy and authenticity of the user, the lack of far-right verified accounts opens up the question of whether there is a greater degree of unverified information spreading in the right-leaning community. We stress, however, that our result is cautionary. A closer investigation is needed to establish if there are other politically driven biases, such as a liberal bias from Twitter as a moderating platform, that may contribute to the underrepresentation of conservative verified users.

While being verified certainly aids visibility and authenticity, users do not need to be verified to be influential. We observed bimodal distributions (U-shaped) in the proportion of users who are influential with respect to their polarity according to three measures of influence: top-most followed, retweeted, and mentioned (Figure 3B-D), indicating that partisan users have more influence in these regards. In particular, far-right users had some of the highest proportion of most-followed users. Far-left users were more likely to be highly retweeted and mentioned, but the far-right also held considerable influence in those regards.

Lastly, we looked at PageRank, a well-known algorithm for measuring node centrality in directed networks [56]. A node with a high PageRank is indicative of high influence and importance. Much like the distribution of verified users, the proportion of users with high PageRank in each polarity decile was correlated with how left-leaning the polarity decile is (Figure 3E), which suggests that left-leaning users hold higher importance and influence. However, this phenomenon may also be an artifact of the much larger left-leaning user base on Twitter.

**Figure 3.** Proportion of users in each decile of predicted political bias scores that are (A) verified, (B) top 5% in the number of followers, (C) top 5% of in-degrees in the retweet network (most retweeted by others), (C) top 5% of in-degrees in the mention network (most mentioned by others), and (E) top 5% in PageRank in the retweet network.

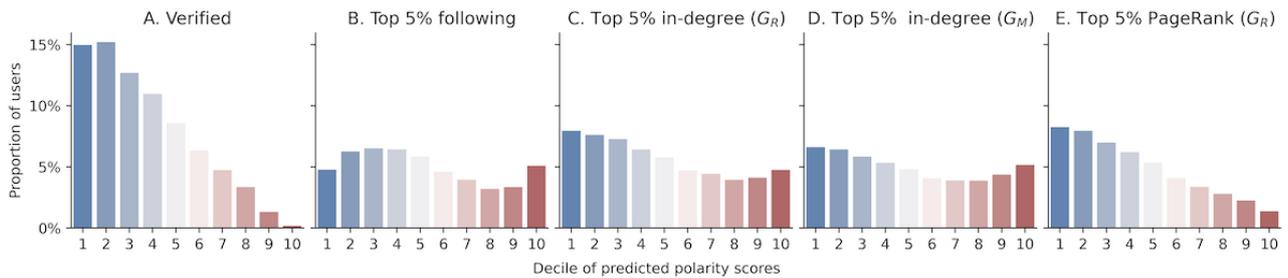

### Echo Chambers

As most influential users are partisan, we questioned the prevalence of echo chambers, if they exist. We began by exploring the partisan relationship between the retweeted and the retweeter, where the latter is considered as the (immediate) audience of the former. Figure 4 plots the proportion of left-leaning, neutral, or right-leaning retweeters for users in each of the 10 deciles of polarity scores, revealing that users on both ends of the political spectrum reached an audience that primarily agrees with their political stance. In fact, the far-left and far-right users had virtually no retweeters from supporters of the opposite party. However, the echo chamber effect was much more prominent on the far-right. About 80% of the audience reached by far-right users were also right-leaning. In comparison, only 40% of the audience reached by far-left users were also left-leaning. There was little difference in the distribution of retweeters between verified and unverified users.

Since the polarized users are mostly preoccupied in their echo chambers, the politically neutral users (Figure 4, green) would serve the important function of bridging the echo chambers and allowing for cross-ideological interactions. Most of them (30%-40%) retweeted from sources that were also neutral, and around 20% of them retweeted from very liberal sources. When it came to broadcasting tweets from far-right users, they behaved similarly to the far-left retweeters: almost no neutral users retweeted from far-right users. Such observations would imply a much stronger flow of communication between the far-left users and neutral users, whereas the far-right users remained in a political bubble.

**Figure 4.** The distribution of left-leaning (bottom 20% of the polarity scores), center (middle 20%), and right-leaning (top 20%) retweeters (y-axis) for users across the polarity score deciles (x-axis). The retweeted users are either verified or not verified.

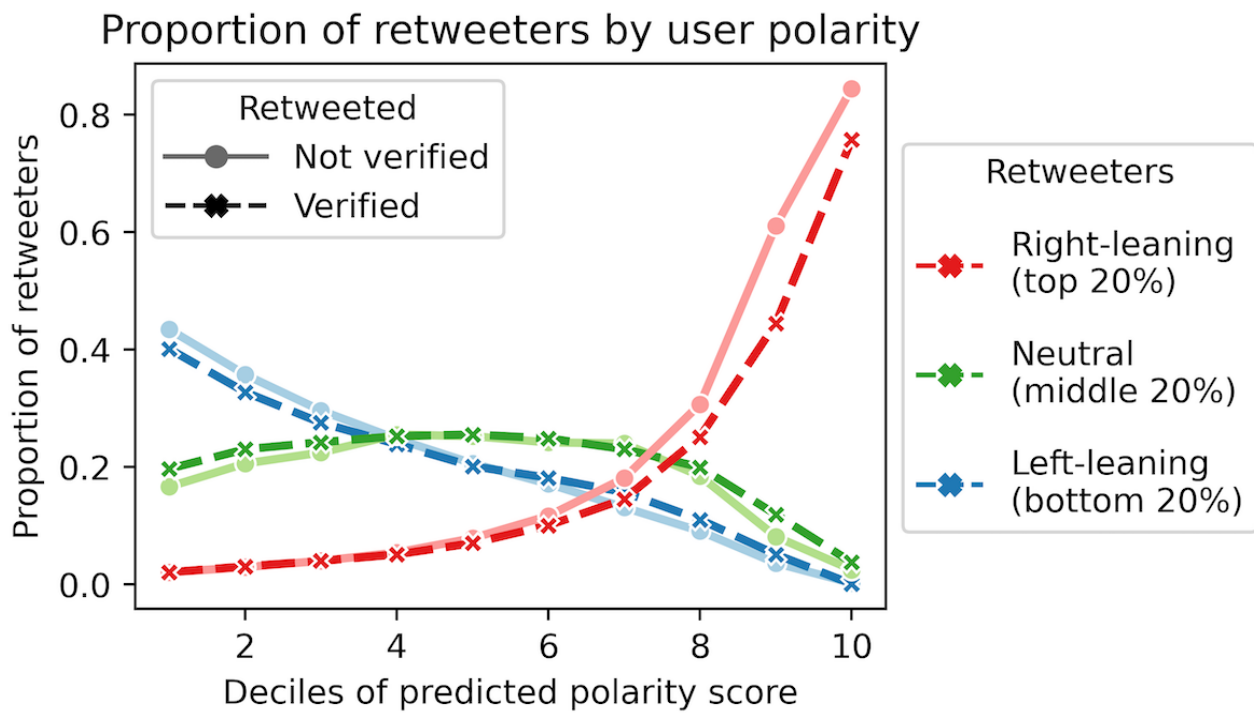

### Random Walk Controversy

Previously, we explored the partisan relationship between users and their immediate audience. To quantify how information is disseminated throughout the Twitter sphere and its relationship with user polarity, we conducted random walks on the graphs to measure the degree of controversy between any two polarity deciles of users. Our method extends the Random Walk Controversy (RWC) score for two partitions [21], which uses random walks to measure the empirical probability of any node from one polarity decile being exposed to information from another.

A walk begins with a given node and recursively visits a random out-neighbor of the node. It terminates when the maximum walk length is reached or if a node previously seen on the walk is revisited. Following Garimella et al [21], we also halted the walk if we reached an authoritative node, which we defined as the top 1000 nodes (≈4%) with the highest in-degree in any polarity decile. By stopping at nodes with high in-degrees, we can capture how likely a node from one polarity decile receives highly endorsed and well-established information from another polarity decile. To quantify the controversy, we measure the RWC from polarity decile $A$ to $B$ by estimating the empirical probability:

$$RWC(A,B) = Pr(\text{start in } A \mid \text{end in } B) \quad (2)$$

The probability is conditional on the walks ending in any partition to control for varying distribution of high-degree vertices in each polarity decile. RWC yields a probability, with a high RWC($A,B$) implying that random walks landing in $B$ started from $A$. Compared to the original work by Garimella et al [21], we simplified the definition of RWC as we did not need to consider the varying number of users in each echo chamber.

We initiated the random walks 10,000 times randomly in each polarity decile for a maximum walk length of 10. The RWC between any two polarity deciles for the retweet and mention networks are visualized in Figure 5. For both networks, the RWC scores were higher along the diagonal, indicating that random walks most likely terminate close to where they originated. Moreover, the intensities of the heatmap visualizations confirmed that there were two separate echo chambers. The right-leaning echo chamber (top-right corner) was much denser and smaller than the left-leaning echo chamber (bottom-left corner). Any walk in the retweet network that originates in polarity deciles 9 and 10 will terminate in polarity deciles 8 to 10 about 80% of the time. In contrast, walks that started in deciles 1 to 7 had a near equal, but overall much smaller, probability of landing in deciles 1 to 7. In essence, users who are right-leaning formed a smaller but stronger echo chamber, while other users formed a larger and more distributed echo chamber.

The RWC scores on the mention network confirmed the presence of the two echo chambers, but the intensities were reduced. Compared to random walks on the retweet network, those on the mention network were much more likely to end far away. As a result, while there were rarely any cross-ideological retweet interactions, there existed a greater degree of direct communication through mentions, likely done to speak to or criticize against the opposing side [8]. We note that, because the RWC scores were highly symmetrical about the diagonals, there was little difference in the cross-ideological interaction between opposite directions of communication flow.

**Figure 5.** The RWC(X,Y) for every pair of polarity deciles X and Y on the retweet (left) and mention (right) networks using equation 2.

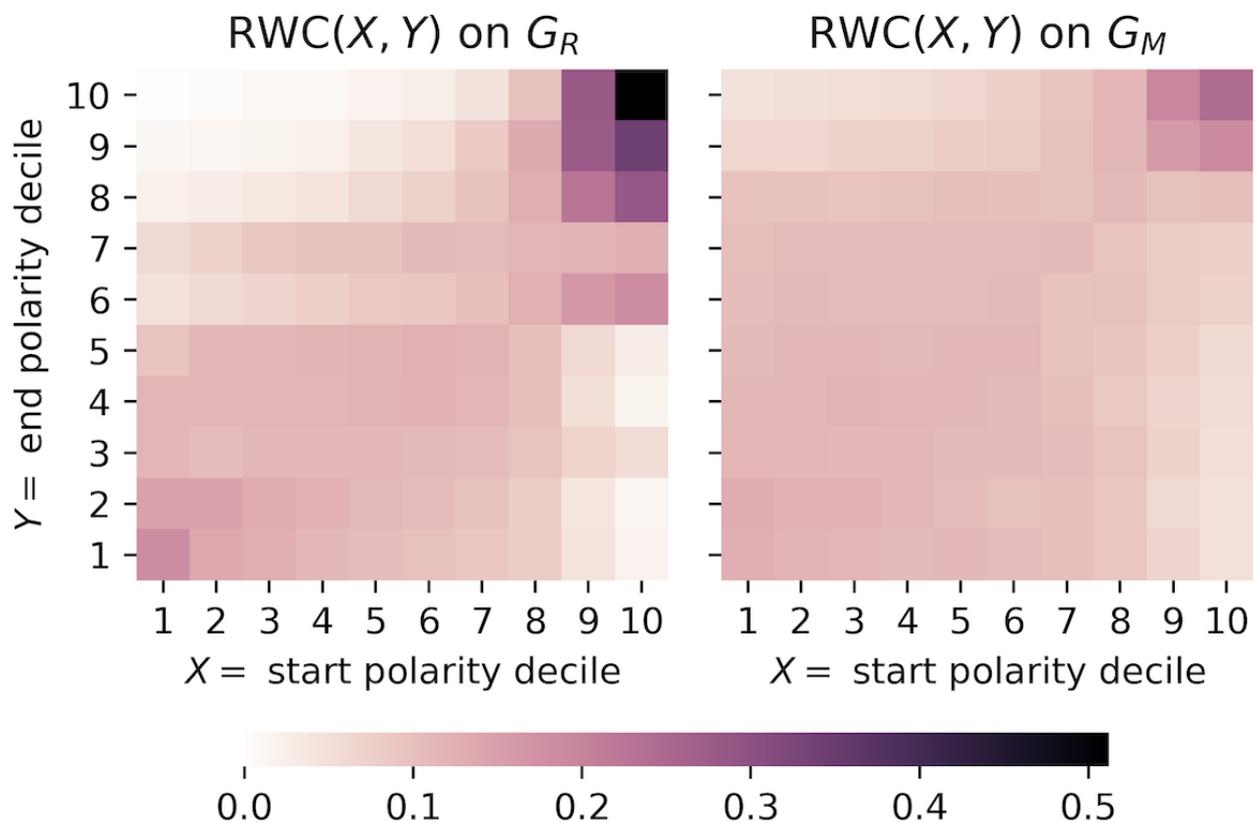

### Popular Users Among the Left and Right

Retweeting is the best indication of active endorsement [22] and is commonly used as the best proxy for gauging popularity and virality on Twitter [57]. Figure 6 shows the most popular users among the left and the right according to the number of left- or right-leaning retweeters they have.

Analyzing the identities of the top-most retweeted users by partisans provides us the first hint at the presence of political echo chambers. There was no overlap between the most retweeted users by the left-leaning and by the right-leaning audience, and they tended to be politically aligned with the polarization of their audience. Almost all users who were most retweeted by left-leaning users were Democratic politicians, liberal-leaning pundits, or journalists working for left-leaning media. Notably, @ProjectLincoln is a political action committee formed by the Republicans to prevent the re-election of the Republican incumbent Donald Trump. Similarly, almost all users who were most retweeted by right-leaning users were Republican politicians, right-leaning pundits, or journalists working for right-leaning media. Despite its username, @Education4Libs is a far-right account promoting QAnon, a far-right conspiracy group. As of January 2021, @Education4Libs had already been banned by Twitter.

These popular users were not only popular among the partisan users but were considerably popular overall, as indicated by the high overall rankings by the number of total retweeters. With a few exceptions, users who were popular among the left were more popular among the general public than users who were popular among the right.

The distribution of the polarity of retweeters of these most popular users revealed another striking observation: the most popular users among the far-right rarely reached an audience that was not also right, whereas those of the far-left reached a much wider audience in terms of polarity. Users who were popular among the far-left hailed the majority of their audience from nonpartisan users (around 75%) and, importantly, drew a sizable proportion of the far-right audience (around 5%). In contrast, users who were popular among the far-right had an audience made up almost exclusively of the far-right (around 80%) and amassed only a negligible amount of the far-left audience.

**Figure 6.** Users with the highest number of retweeters from left- and right-leaning users. The bar plots show the distribution of their unique retweeters by political leaning. Users are also ranked by their total number of retweeters (ie, "#1 @realDonaldTrump" means that @realDonaldTrump has the most retweeters). Numbers appended to the end of the bars show the total number of retweeters.

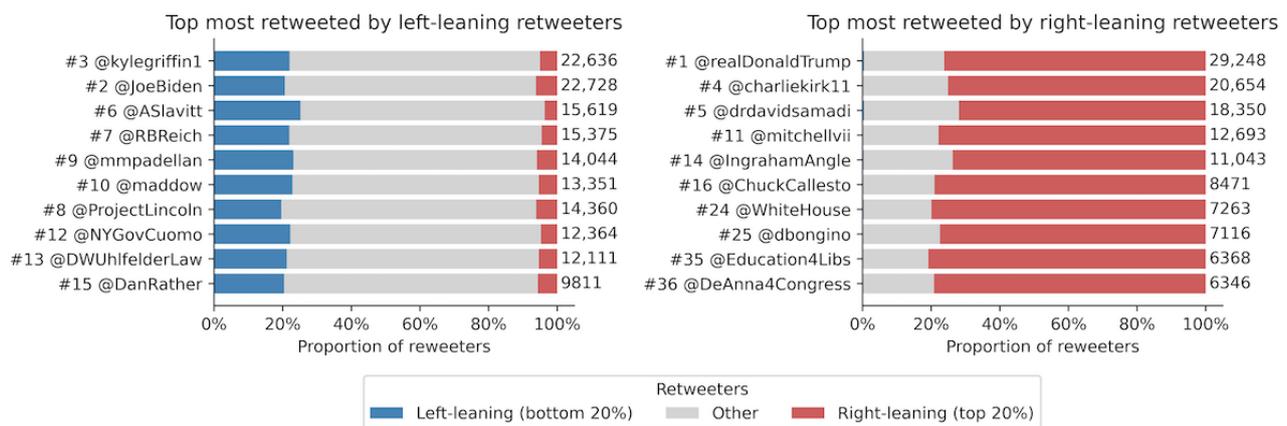

## Discussion

In this paper, we study the extent of echo chambers and political polarization in COVID-19 conversations on Twitter in the United States. We proposed Retweet-BERT, a model that leverages user profile descriptions and retweet interactions to effectively and accurately measure the degree and direction of polarization. Applying Retweet-BERT, we provided insightful characterizations of partisan users and the echo chambers in the Twitter sphere to address our research questions.

### RQ1: What Are the Roles of Partisan Users on Social Media in Spreading COVID-19 Information? How Polarized Are the Most Influential Users?

From characterizing partisan users, we found that right-leaning users stand out as being more vocal, more active, and more impactful than their left-leaning counterparts.

Our finding that many influential users are partisan suggests that online prominence is linked with partisanship. This result is in line with previous literature on the "price of bipartisanship," which is that bipartisan users must forgo their online influence if they expose information from both sides [28]. In another simulated study, Garibay et al [58] showed that polarization can allow influential users to maintain their influence. Consequently, an important implication is that users may be incentivized to capitalize on their partisanship to maintain or increase their online popularity, thereby further driving polarization. Information distributed by highly polarized yet influential users can reinforce political predispositions that already exist, and any polarized misinformation spread by influencers risks being amplified.

### RQ2: Do Echo Chambers Exist? If Yes, What Are the Echo Chambers and How Do They Compare?

Though COVID-19 is a matter of public health, we discovered strong evidence of political echo chambers on this topic on both ends of the political spectrum, but particularly within the right-leaning community. Right-leaning users were almost exclusively retweeted by users who were also right-leaning, whereas the left-leaning and neutral users had a more proportionate distribution of retweeter polarity. From random walk simulations, we found that information rarely traveled in or out of the right-leaning echo chamber, forming a small yet intense political bubble. In contrast, far-left and nonpartisan users were much more receptive to information from each other. Comparing users who are popular among the far-left and the far-right, we revealed that users who were popular among the right were *only* popular among the right, whereas users who were popular among the left were also popular among all users.

## Implications

Despite Twitter's laudable recent efforts in fighting misinformation and promoting fact checking [59], we shed light on the fact that communication is not just falsely manipulated, but also hindered, by communication bubbles segregated by partisanship. It is imperative that we not only dispute misinformation but also relay true information to all users. As we have shown, outside information is extremely difficult to get through to the right-leaning echo chamber, which could present unique challenges for public figures and health officials outside this echo chamber to effectively communicate information. Existing research suggests that right-leaning users are more susceptible to antiscience narratives, misinformation, and conspiracy theories [2,3,17,18], which given the echo chambers they are situated in can worsen with time. Our work has implications in helping officials develop public health campaigns, encourage safe practices, and combat vaccine hesitancy effectively for different partisan audiences.

## Future Direction

Though the question of whether social media platforms *should* moderate polarization is debated, we note that *how* they can do so remains an open problem. It is unclear how much of the current polarization is attributed to users' selective exposure versus the platform's recommendation algorithm. Moreover, whether users are even aware that they are in an echo chamber, and how much conscious decision is being made by the users to combat that, remains to be studied in future work.

Another future avenue of research could focus on studying how misinformation travels in different echo chambers. Since our study highlights that there is an alarmingly small number of far-right verified users, and given that verified users are typically believed to share legitimate and authentic information, further research is required to establish if the right-leaning echo chamber is at greater risk of being exposed to false information from unverified users. A detailed content analysis of tweets can reveal if there are significant disparities in the narratives shared by left- and right-leaning users. Crucially, our work provides a basis for more in-depth analyses on how and what kind of misinformation is spread in both echo chambers.

## Limitations

There are several limitations regarding this work. First, we cannot exclude any data bias. The list of keywords was manually constructed, and the tweets collected are only a sample of all possible tweets containing these keywords. Since the data was collected based on keywords strictly related to COVID-19, we only gathered data that is relevant to the virus and not tainted by political commentary. Therefore, the data provides us a natural setting to study the polarization of COVID-19 discourse on Twitter.

Second, our study hinges on the fact that retweets imply endorsement, which may be an oversimplification. To reduce noisy, isolated retweet interactions, we considered only retweets that have occurred at least twice between any two users.

Finally, our political detection model was built on a weakly supervised labeling of users using politically relevant hashtags and the polarization of news media as the sources of ground-truth. We took a conservative approach and only seeded users who explicitly used politicized hashtags in their profile or had repeatedly interacted with polarized new sources.

## Conflicts of Interest

None declared.

## Multimedia Appendix 1

Supplementary material.
[[PDF File (Adobe PDF File), 93 KB](#)-[Multimedia Appendix 1](#)]

XSL•FO
RenderX

## Abbreviations

**AUC:** area under the receiver operating characteristic curve
**RQ:** research question
**RWC:** Random Walk Controversy